\documentclass[conference]{IEEEtran}
\IEEEoverridecommandlockouts
\usepackage{cite}
\usepackage{amsmath,amssymb,amsfonts}
\usepackage{algorithmic}
\usepackage{graphicx}
\usepackage{textcomp}
\usepackage{booktabs} 
\usepackage[a4paper, total={184mm,239mm}]{geometry}
\def\BibTeX{{\rm B\kern-.05em{\sc i\kern-.025em b}\kern-.08em
    T\kern-.1667em\lower.7ex\hbox{E}\kern-.125emX}}
\begin{document}

\title{\fontsize{23.96pt}{\baselineskip}\selectfont Gem5-AcceSys: Enabling System-Level Exploration of Standard Interconnects for Novel Accelerators\\}

\author{
	\IEEEauthorblockN{
	Qunyou Liu\IEEEauthorrefmark{1},
    Marina Zapater\IEEEauthorrefmark{2},
    David Atienza\IEEEauthorrefmark{1},  \\
	}			  
    \IEEEauthorblockA{\IEEEauthorrefmark{1}\textit{Embedded Systems Laboratory (ESL), EPFL, Switzerland, }\textit{qunyou.liu@epfl.ch, david.atienza@epfl.ch}}    
    \IEEEauthorblockA{\IEEEauthorrefmark{2}\textit{University of Applied Sciences and Arts Western Switzerland (HES-SO), Switzerland, }\textit{marina.zapater@heig-vd.ch}}
} 
\IEEEaftertitletext{\vspace{-1.0\baselineskip}} 

\maketitle
\begin{abstract}
The growing demand for efficient, high-performance processing in machine learning (ML) and image processing has made hardware accelerators, such as GPUs and Data Streaming Accelerators (DSAs), increasingly essential. These accelerators enhance ML and image processing tasks by offloading computation from the CPU to dedicated hardware. These accelerators rely on interconnects for efficient data transfer, making interconnect design crucial for system-level performance. This paper introduces Gem5-AcceSys, an innovative framework for system-level exploration of standard interconnects and configurable memory hierarchies. Using a matrix multiplication accelerator tailored for transformer workloads as a case study, we evaluate PCIe performance across diverse memory types (DDR4, DDR5, GDDR6, HBM2) and configurations, including host-side and device-side memory. Our findings demonstrate that optimized interconnects can achieve up to 80\% of device-side memory performance and, in some scenarios, even surpass it. These results offer actionable insights for system architects, enabling a balanced approach to performance and cost in next-generation accelerator design.

\end{abstract}

\vspace{-0.25cm}
\section{Introduction}
\vspace{-0.15cm}
Transformer models, such as Vision Transformer (ViT)~\cite{Dosovitskiy2021}, have revolutionized machine learning (ML) and natural language processing (NLP), setting new benchmarks in various tasks~\cite{Devlin2019}. Their growing adoption in real-time applications demands efficient and scalable hardware architectures to meet rising computational needs. Systolic arrays have shown promise in accelerating these tasks, offering parallel data flow and high computational throughput ideal for matrix multiplication, central to both traditional ML and transformer-based models, such as Google's Tensor Processing Unit~\cite{Jouppi2017} and TiC-SAT~\cite{Amirshahi2023}. However, most studies on systolic-array-based architectures primarily focus on hardware-level evaluations, often neglecting the broader system-level context. This leaves a gap in understanding how the interconnects and memory hierarchy, which are vital components in moving data between processors and memory, impact the overall system performance. While some system-level evaluations exist, they often lack the use of standard interconnects such as PCIe, and heterogeneous memory hierarchy such as non-uniform memory access (NUMA), reducing their practicality for real-world applications.

To bridge this gap, we introduce Gem5-AcceSys, a novel framework enabling system-level exploration of interconnects and memory hierarchies in accelerators. Using a matrix multiplication accelerator tailored for transformer workloads as a case study, our design framework incorporates essential components such as PCIe~\cite{Ajanovic2009},~\cite{Vasa2020}, NUMA architectures, and configurable memory hierarchies. This integration provides a practical platform for evaluating real-world performance under various configurations. Specifically, we assess the impact of standard interconnects (PCIe) and diverse memory types, including DDR3/4/5, GDDR6, and HBM2, on system efficiency. By emphasizing system-level architecture, our work provides a comprehensive understanding of how interconnects and memory hierarchies influence transformer's performance and how general matrix multiplication (GEMM) and Non-GEMM impact the system performance with the overhead introduced by standard interconnects. This research offers actionable insights for designing scalable and cost-effective accelerator systems tailored to modern ML workloads. More specifically, our contributions are as follows:
\begin{enumerate}
    \item \textbf{Framework for Interconnect Exploration}: We present the framework, Gem5-AcceSys, to support PCIe interconnects, NUMA architecture, and processing near memory based on the Gem5 simulator, enabling system-level co-design with features like device-side memory, local buffer, SMMU, DMA, and RTL-based accelerators.
    
    \item \textbf{PCIe Bandwidth impact on Performance Study}: We investigate the impact of varying PCIe bandwidth by adjusting the number of lanes and lane speeds for the target GEMM workload. Additionally, we vary the packet request size from the perspective of the accelerator to identify the optimal packet size for the design.
    
    \item \textbf{Memory Access and Address Translation Study}: We analyze the impact of memory type and location on system performance from a system-level perspective. Additionally, we investigate address translation overhead for Transformer workloads, demonstrating the effects of virtual address spaces on accelerator efficiency.

    \item \textbf{Impact of Non-GEMM and GEMM Workload Analysis}: We conduct a detailed runtime analysis of Transformer workloads, dividing them into two components: GEMM and Non-GEMM. To optimize performance, we profiled these workloads separately and identified thresholds to determine when device-side memory should be utilized.
\end{enumerate}

\vspace{-0.5cm}
\section{State-of-the-art}
\vspace{-0.15cm}
Several open-source simulators have been developed to aid in the design and integration of specialized hardware accelerators, notably Gem5-Aladdin~\cite{Shao2016}, Gem5-Salam~\cite{Rogers2020}, Gem5-RTL~\cite{Lopez2021}, and Gem5-X~\cite{Qureshi2021}. These frameworks provide valuable tools for early-stage design exploration and detailed evaluations of accelerators within system simulations. Gem5-Aladdin integrates accelerator modeling with system simulation, enabling pre-RTL design analysis. Gem5-Salam employs LLVM-based methodologies to model and estimate the performance and area of custom accelerators. Gem5-RTL incorporates RTL designs for detailed performance assessments, while Gem5-X supports simulations of many-core heterogeneous systems with advanced memory technologies.
However, these simulators have limitations that hinder their applicability for comprehensive system-level co-design. They rely on simplistic interconnects (e.g., basic buses, shared memory), lacking support for standard interfaces like PCIe. Memory hierarchy support is limited, with features like Non-Uniform Memory Access (NUMA) and processing near memory (PNM) absent~\cite{Khan2024}. High simulation overhead due to RTL integration, as seen in frameworks like Gem5-RTL, impedes rapid prototyping. Scalability issues arise when accurately simulating diverse accelerators within large-scale systems. Additionally, inadequate system integration is evident from the lack of features like Direct Memory Access (DMA) and System Memory Management Unit (SMMU) for efficient data handling~\cite{Whitham2010},~\cite{Paraskevas2020}. These limitations highlight the need for a comprehensive framework that supports standard interconnects, complex memory hierarchies, and realistic full-system interactions. To address this gap, we propose \textbf{Gem5-AcceSys} to enable detailed system-level co-design of accelerators.
\vspace{-0.25cm}
\section{Design framework}
\vspace{-0.25cm}

\subsection{Overall Architecture and Interface Integration}
\begin{figure}[h]
\vspace{-0.2cm}
  \centering
  \includegraphics[width=0.8\linewidth]{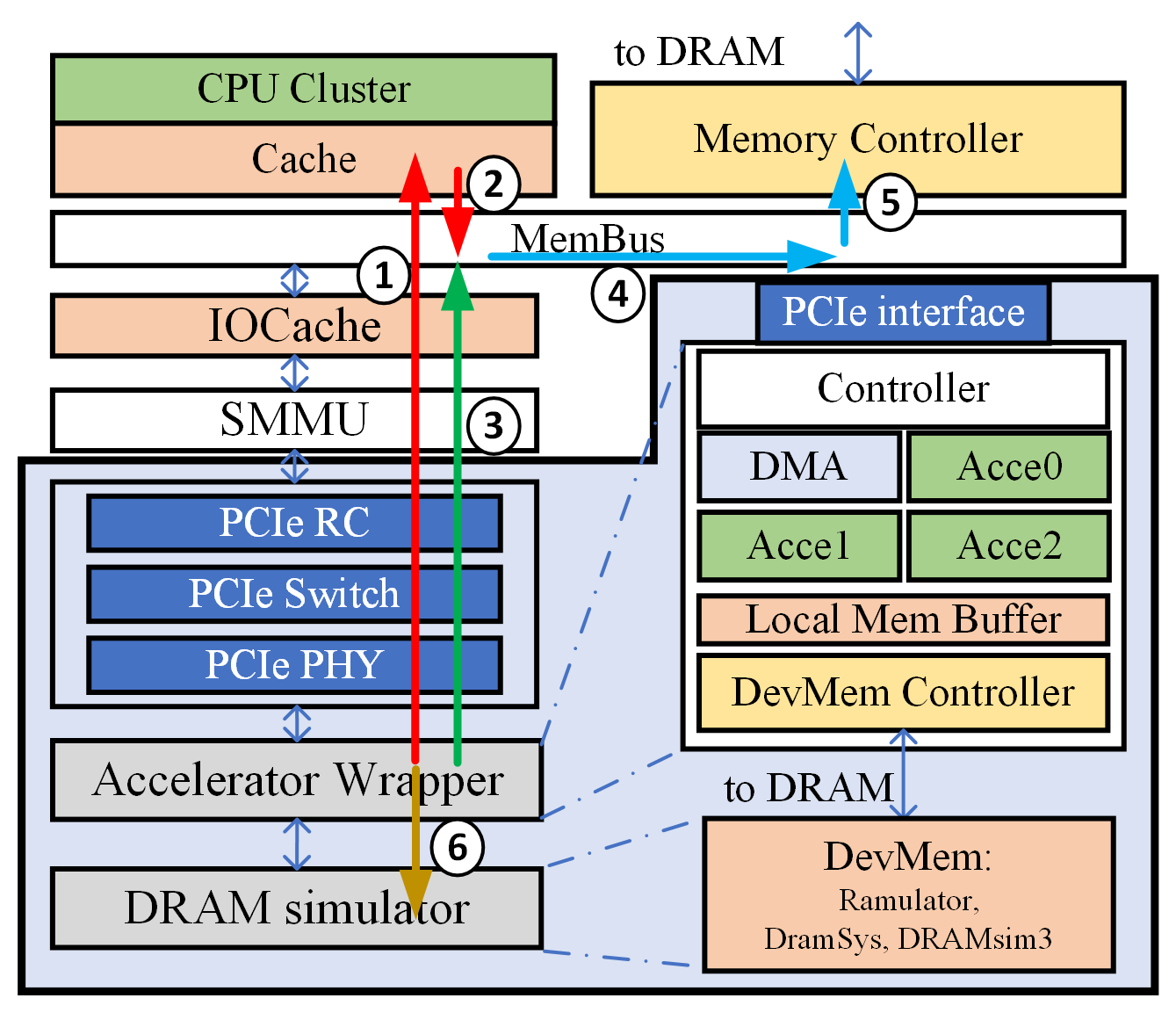}
  \vspace{-0.25cm}
  \caption{Design Framework Architecture}
  \label{fig:designFrame}
  \vspace{-0.4cm}
\end{figure}

\vspace{-0.15cm}
Fig.~\ref{fig:designFrame} illustrates our enhanced design framework, with newly added components highlighted in light blue, significantly expanding the Gem5 simulator's capabilities for contemporary system design. This framework integrates essential functionalities, including PCIe interconnects, DMA, SMMU, configurable sub-memory architecture, and RTL-based accelerator support. The system is organized into two primary sections: the CPU cluster with its cache and the accelerator system interfaced with the memory bus (MemBus) through PCIe components.

The CPU cluster, comprising one or more CPUs and associated caches, connects to the main memory (DRAM) via a memory controller. Data transfers between the CPU and memory are facilitated by a MemBus. PCIe integration provides a realistic interface for peripheral device simulations, surpassing the limitations of conventional memory bus latency models. The DMA feature further enhances system efficiency by enabling direct memory transfers, reducing the data movement burden on the accelerator and simplifying its design. The SMMU enables virtual-to-physical address translation, streamlining driver development and improving memory mapping efficiency for accelerators~\cite{Whitham2010},~\cite{Paraskevas2020}. A configurable sub-memory system supports the optimization of memory configurations crucial for co-designing accelerator systems. This includes device-side memory with local buffers and host-side memory configurations to evaluate different system trade-offs. Key PCIe components include:

\begin{itemize}
    \item \textbf{PCIe RC (Root Complex)}: The CPU endpoint that manages data and commands on the PCIe bus.
    \item \textbf{PCIe PHY (Physical Layer)}: Connects the RC to devices, facilitating communication via the PCIe Switch and Link.
    \item \textbf{PCIe Switch}: Routes traffic among PCIe devices, supporting multiple connections and enhancing scalability.
\end{itemize}

By positioning the SMMU between the MemBus and PCIe components, the framework enhances memory protection and address translation capabilities, enabling efficient use of virtual address spaces.
\vspace{-0.25cm}
\subsection{Accelerator Wrapper}




The Accelerator Wrapper contains logic and interfaces for the hardware accelerator, including a PCIe interface that connects to the accelerator's controller. This controller manages data flows between the accelerator and system. Adjacent to this, the single accelerator or accelerator cluster, either RTL-based or C++-based, integrates with the system using Verilator~\cite{Chi2022} to convert RTL code to C++, compiling it into an executable that runs as a child process with shared memory system calls. A DMA block within the controller enables direct memory access, bypassing the CPU for higher performance. A Local Mem Buffer provides quick, temporary computation storage, and a Device Memory (DevMem) Controller oversees data transfers between the accelerator and device memory, boosting efficiency. The accelerator also interfaces with a configurable memory hierarchy.

\vspace{-0.25cm}
\subsection{Memory Hierarchy and Subsystem Performance Analysis}
Our design framework incorporates a configurable memory hierarchy, including host-side and device-side memories, along with a last-level cache and a device-side cache, supporting comprehensive design space exploration. It interfaces with DRAM models like DRAMsim3~\cite{Li2020}, Ramulator2~\cite{Luo2023}, and Dramsys5~\cite{Steiner2022} for accurate DRAM timing and power statistics. The framework offers three memory access methods: direct cache (DC) access, direct memory (DM) access and DevMem access. Fig.~\ref{fig:designFrame} illustrates DC and DM modes. In DC mode, data requests from the accelerator are directed to the cache hierarchy. Cache hits retrieve data immediately, reducing latency, while cache misses access host-side memory via the MemBus (arrow 2) and memory controller (arrows 4 and 5), adding latency. This mode enables detailed analysis of cache effects on performance and allows adjustments to cache size and latency. To support this, we modified the simulator to implement a cache coherency model between the accelerator's cache and the CPU cache. In DM mode, requests bypass the cache and go straight to main memory via arrows 3 and 5, minimizing latency by eliminating cache checks. This method requires software management of data coherency since it bypasses the cache system. Arrows 6 bypasses the whole PCIe system to access device-side memory avoid the data movement overhead introduce by PCIe system. By leveraging configurable memory hierarchies and flexible access methods, our design framework enables the selection of optimal memory configurations to enhance accelerator performance. Additional insights into the impact of memory hierarchy on performance are discussed in Section~\ref{sec:expeandres}.

\vspace{-0.25cm}
\subsection{Comparison with Existing Simulators}
To highlight the unique features of Gem5-AcceSys and how it extends beyond existing Gem5-based simulators, we provide a concise comparison in Table~\ref{tab:simulator_comparison}. This table summarizes key features such as accelerator (Acce) design level support, interconnect capabilities, address translation, memory simulation, kernel driver support, DMA capabilities, device-side memory support, simulation scope, and accelerator process models.

\begin{table*}[htbp]
\vspace{-0.5cm}
\centering
\caption{Comparison of Gem5-based frameworks for hardware accelerator simulation}
\label{tab:simulator_comparison}
\begin{tabular}{lccccc}
\toprule
\textbf{Feature} & \textbf{Gem5-Aladdin} & \textbf{Gem5-SALAM} & \textbf{Gem5-RTL} & \textbf{Gem5-X} & \textbf{Gem5-AcceSys} \\
\midrule
\textbf{Acce Design Level} & C++ & LLVM IR & RTL & C++ & C++, RTL \\
\textbf{Interconnect} & Basic buses & Basic buses & Basic buses & Basic buses & Basic buses, PCIe \\
\textbf{Acce Address Translation} & Yes & No & No & No & Yes (SMMU) \\
\textbf{External Memory simulator} & No & No & No & No & Ramulator/DRAMsys \\
\textbf{Kernel Driver Support} & No & No & No & Limited & Yes \\
\textbf{Multi-Channel DMA} & Yes & No & No & No & Yes \\
\textbf{Device-Side Memory} & No & No & No & Yes & Yes \\
\textbf{Full-System Simulation} & Yes & Bare-metal & Yes & Yes & Yes \\
\textbf{Acce Process Model} & Integrated & Integrated & Integrated & Integrated & Child process (Multi-threaded) \\
\bottomrule
\end{tabular}
\end{table*}
As shown in Table~\ref{tab:simulator_comparison}, Gem5-AcceSys provides comprehensive support for features essential in modern accelerator design, addressing limitations present in existing simulators. This includes support for standard interconnects like PCIe, accelerator address translation via SMMU, integration with external memory simulators, kernel driver support, multi-channel DMA, and device-side memory.
\vspace{-0.25cm}
\section{Experimental setup}
\vspace{-0.25cm}
To demonstrate the capabilities of our design framework and analyze the impact of standard interconnects and configurable memory architectures on transformer workloads, we conduct experiments using the DC access method, with configurations and parameters detailed in Table~\ref{tab:system_configuration}.
\vspace{-0.1cm}
\begin{table}[h]
\vspace{-0.5cm}
\caption{System Configuration}
\centering
\begin{tabular}{lc}
\hline
\textbf{Component} & \textbf{Specification}\\
\hline
CPU & ARM, 1 GHz \\
Data Cache & 64 kB\\
Instruction Cache & 32 kB\\
Last Level Cache & 2 MB \\
IOCache & 32 kB \\
Memory & DDR3\_1600\_8x8, 4 GB \\
PCIe Link & Version 2.0, 4 Gb/s, 4 Lanes \\
PCIe RootComplex & 150ns Latency \\
PCIe Switch & 50ns Latency \\
\hline
\end{tabular}
\label{tab:system_configuration}
\vspace{-0.5cm}
\end{table}
\vspace{-0.25cm}
\subsection{Hardware Accelerator Design}

We use a Systolic Array (SA) accelerator named MatrixFlow~\cite{Liu2025}. MatrixFlow is a loosely-coupled SA specifically optimized for transformer models. It uses a new matrix multiplication technique alongside an optimized data structure, which significantly enhances data streaming efficiency and reduces memory overhead. MatrixFlow contains 16x16 multiply–accumulate units, and uses data in integer format as input and output. Through a combination of hardware-software co-design, including the integration of PCIe and DMA for efficient data handling, MatrixFlow achieves remarkable speed-ups of up to 400x compared to a single-threaded CPU system, and demonstrates substantial performance improvements over both loosely- and tightly-coupled accelerators in the state-of-the-art.
\vspace{-0.5cm}
\subsection{Workload}
First, we select GEMM as a workload because it represents a general and widely used computational pattern, showcasing the versatility of our design framework in evaluating performance and analyzing bottlenecks. Next, to further illustrate our framework's capabilities in co-optimizing complex systems, we use the Vision Transformer (ViT)~\cite{Dosovitskiy2021} as a case study. We evaluate workloads from the ViT\_base, ViT\_large, and ViT\_huge models, which have hidden dimensions of 768, 1024, and 1280, respectively, and utilize 12 or 16 attention heads.
\vspace{-0.25cm}
\section{Performance Evaluation}
\vspace{-0.25cm}
\label{sec:expeandres}

To evaluate the performance of our design framework, we focus first on GEMMs and analyze  its performance under various configurations. Based on the insights gained from this analysis, we narrow-down the set of interesting configurations and focus on the entire transformer.
\vspace{-0.25cm}
\subsection{Performance Bounds and Roofline Model}
We begin by analyzing the roofline model of our accelerator system, focusing on GEMMs with dimensions 1024. To isolate the effects of computation on performance, we fix the PCIe bandwidth at 8 GB/s and vary the computation time of the systolic array within the Gem5 framework.

\begin{figure}[h]
\vspace{-0.5cm}
  \centering
  \includegraphics[width=\linewidth]{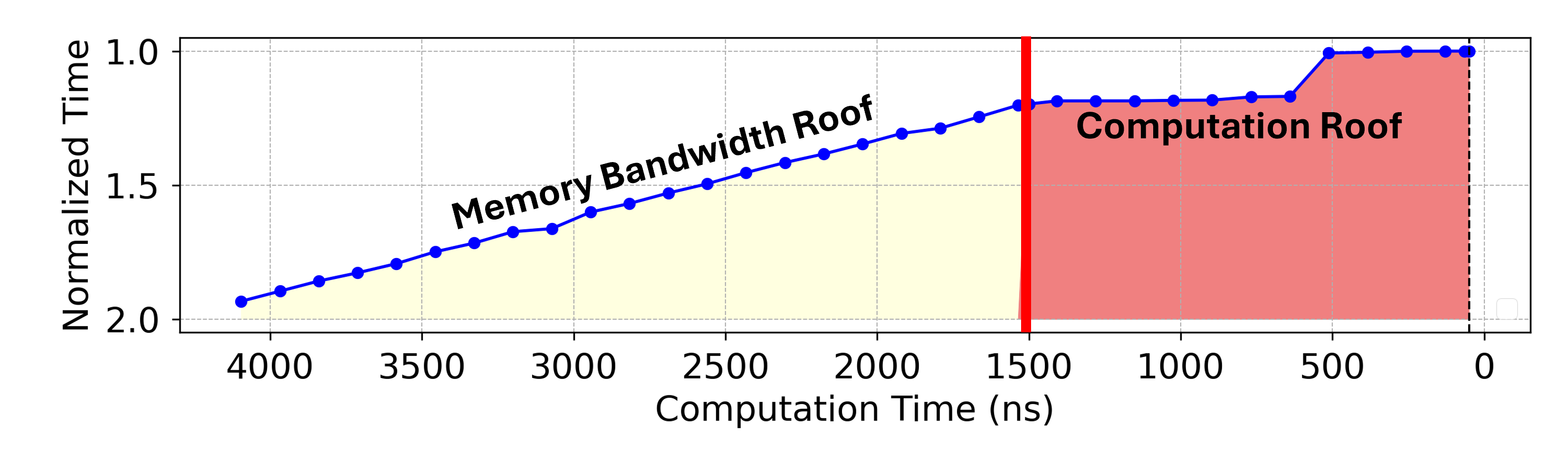}
  \vspace{-0.8cm}
  \caption{Roofline Model of the Accelerator System}
  \label{fig:roofline}
  \vspace{-0.35cm}
\end{figure}

Fig.~\ref{fig:roofline} illustrates the roofline model of the proposed accelerator. The x-axis represents computation time (ns), while the y-axis shows the normalized execution time. For computation times exceeding 1500 ns, the system operates in the \textbf{memory-bound} region, where performance improves linearly as computation time decreases. This indicates that performance is limited by memory bandwidth or PCIe data transfer speed.
As computation time drops below 1500 ns, the system transitions to the \textbf{computation-bound} region, as shown by the plateau in performance. Here, further improvements in memory bandwidth have no significant effect because the computation speed of the systolic array becomes the bottleneck. This transition, marked by the red line, highlights the shift in limiting factors. To better understand these performance bounds, we proceed with a detailed analysis of GEMM workloads.
\vspace{-0.25cm}
\subsection{GEMMs}
\label{sec:GEMMres}
We start by analyzing the impact of the following factors:
\begin{itemize}
    \item \textbf{PCIe Link:} Analyzing how PCIe bandwidth, latency , and packet sizes affect the data transfer efficiency and overall computation speed.
    \item \textbf{Memory Type, location, latency and bandwidth:} Assessing performance variations between device-side and host-side memory, under different memory technologies including  DDR4, LPDDR, GDDR, and HBM.
    \item \textbf{Address Translation:} Analysing the overhead of Address Translation wrt matrix size and their effects on computational delays and resource utilization.
\end{itemize}

\subsubsection{PCIe Link Performance Analysis}

\begin{figure}[h]
\vspace{-0.25cm}
  \centering
  \includegraphics[width=\linewidth]{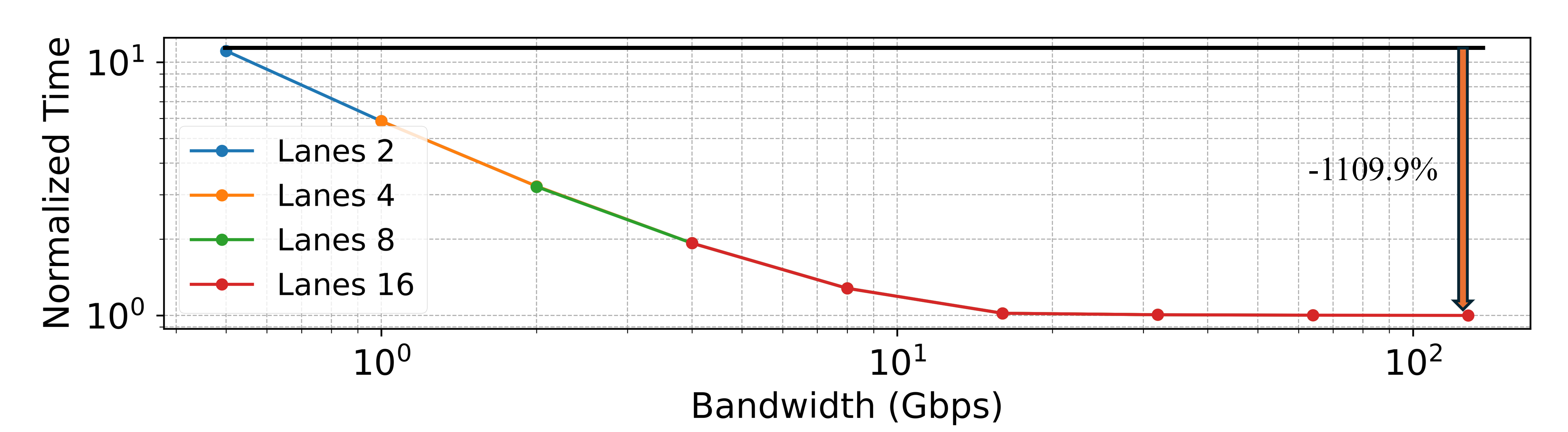}
  \vspace{-0.8cm}
  \caption{Performance (Execution time) for Matrix Size 2048 under varying per-lane bandwidth and number of lanes}
  \vspace{-0.25cm}
  \label{fig:GEMM_bandwidth}
\end{figure}

\paragraph{Bandwidth Sweeping}
To evaluate the impact of PCIe bandwidth on GEMM performance, we vary the number of lanes (2, 4, 8, 16) and their speeds (2 Gbps to 64 Gbps), using 2048x2048 matrices on the systolic array accelerator.  As depicted in Fig.~\ref{fig:GEMM_bandwidth}, increasing bandwidth consistently reduces execution time, with performance scaling until the system transitions from memory-bound to compute-bound at 16 lanes. Notably, bandwidth impacts performance significantly, with the highest bandwidth outperforming the lowest by up to 1109.9\%. This highlights the effectiveness of our framework in identifying optimal bandwidth configurations for balanced performance and resource efficiency.


\textbf{\textit{Key Takeaway \#1:} PCIe bandwidth significantly impacts accelerator performance, particularly in memory-bound regions. However, as systems become compute-bound, the benefits of additional bandwidth diminish.}

\begin{figure}[h]
  \centering
  \includegraphics[width=\linewidth]{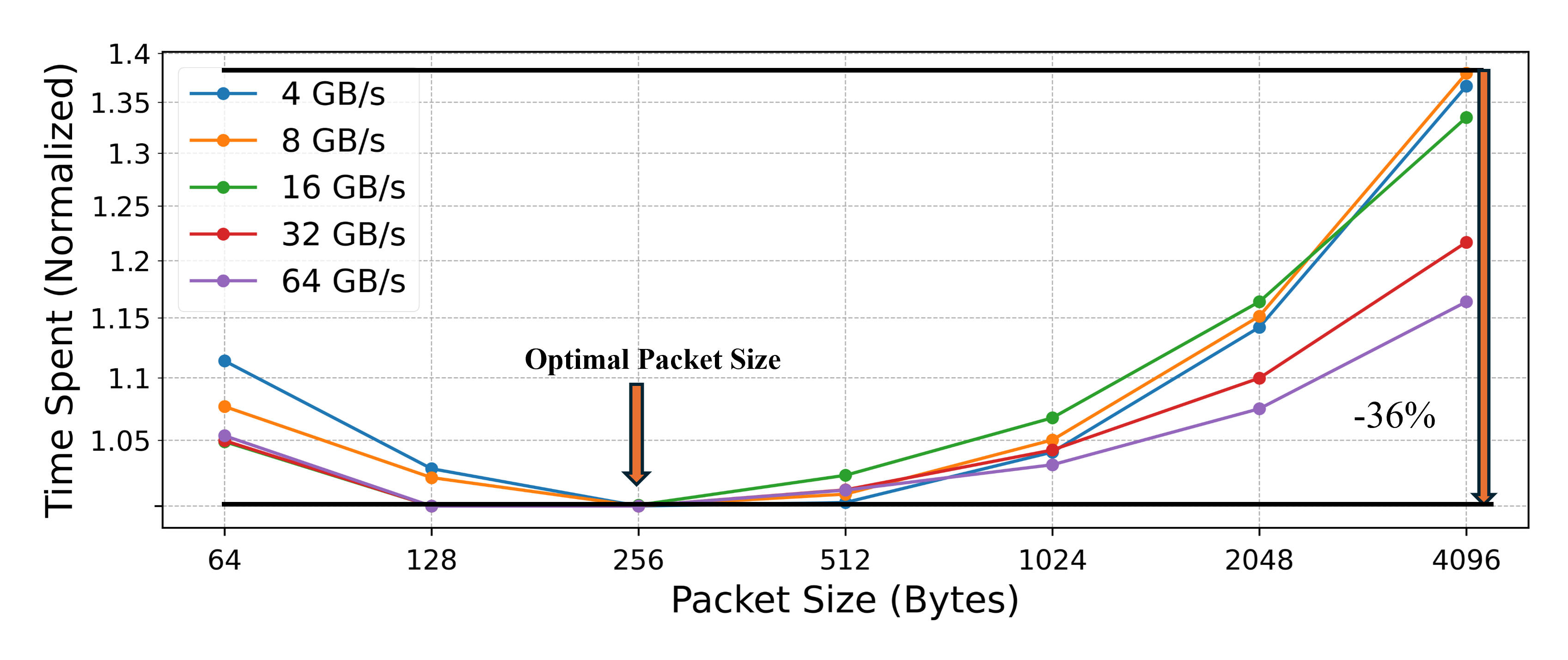}
  \vspace{-0.8cm}
  \caption{Execution Time under different packet sizes for different PCIe bandwidth.}
  \vspace{-0.25cm}
  \label{fig:GEMM_packet}
  \vspace{-0.25cm}
\end{figure}

\paragraph{Packet Size Sweeping}
We configure the PCIe Link at 4GB/s, 8GB/s, 16GB/s, 32GB/s, and 64GB/s, analyzing the impact of request packet sizes from accelerator ranging from 64 bytes to 4096 bytes. As shown in Fig.~\ref{fig:GEMM_packet}, varying packet sizes result in differing processing times, even reach 36\%. Initially, execution time decreases as packet size increases from 64 bytes, reaching a minimum around 256 bytes, reflecting efficiency gains at moderate sizes. Beyond this point, execution time increases with larger packet sizes up to 4096 bytes, forming a convex curve that highlights a non-linear relationship: both very small and very large packet sizes are less efficient. This behavior stems from the PCIe hierarchy, where packets must pass through the RootComplex and Switch. Larger packets disrupt the pipeline, causing stalls at each component before data reaches the accelerator.

\textbf{\textit{Key Takeaway \#2:} Packet size significantly affects execution time with a convex trend: 64-byte packets incur 12\% overhead, and 4096-byte packets 36\%, relative to the optimal 256-byte size.}

\subsubsection{Memory performance analysis}
To examine the performance impact of different memory types and locations, we use ramulator2 as our backend DRAM model, and test DDR3, DDR4, HBM, and GDDR5 with the bandwidth and data rate configurations detailed in Table~\ref{tab:mem_config}.

\begin{table}[h]
\vspace{-0.5cm}
\caption{Memory Configuration}
\centering
\begin{tabular}{lcccc}
\hline
\textbf{Component} & \textbf{Channel} & \textbf{Data width} & \textbf{Bandwidth} & \textbf{Data Rate} \\
\hline
DDR3 &  1 &  64 & 12.8 GB/s & 1600 MT/s \\
DDR4 &  1 &  64 & 19.2 GB/s & 2400 MT/s \\
DDR5 &  2 &  32 & 25.6 GB/s & 3200 MT/s \\
HBM2 & 2 &  128 & 64 GB/s & 2000 MT/s \\
GDDR6 & 2 &  64 & 32 GB/s & 2000 MT/s \\
\hline
\end{tabular}
\label{tab:mem_config}
\end{table}
\vspace{-0.5cm}

\paragraph{Device-side vs Host-side Memory and Memory Type}

\begin{figure}[h]
  \centering
  \includegraphics[width=0.9\linewidth]{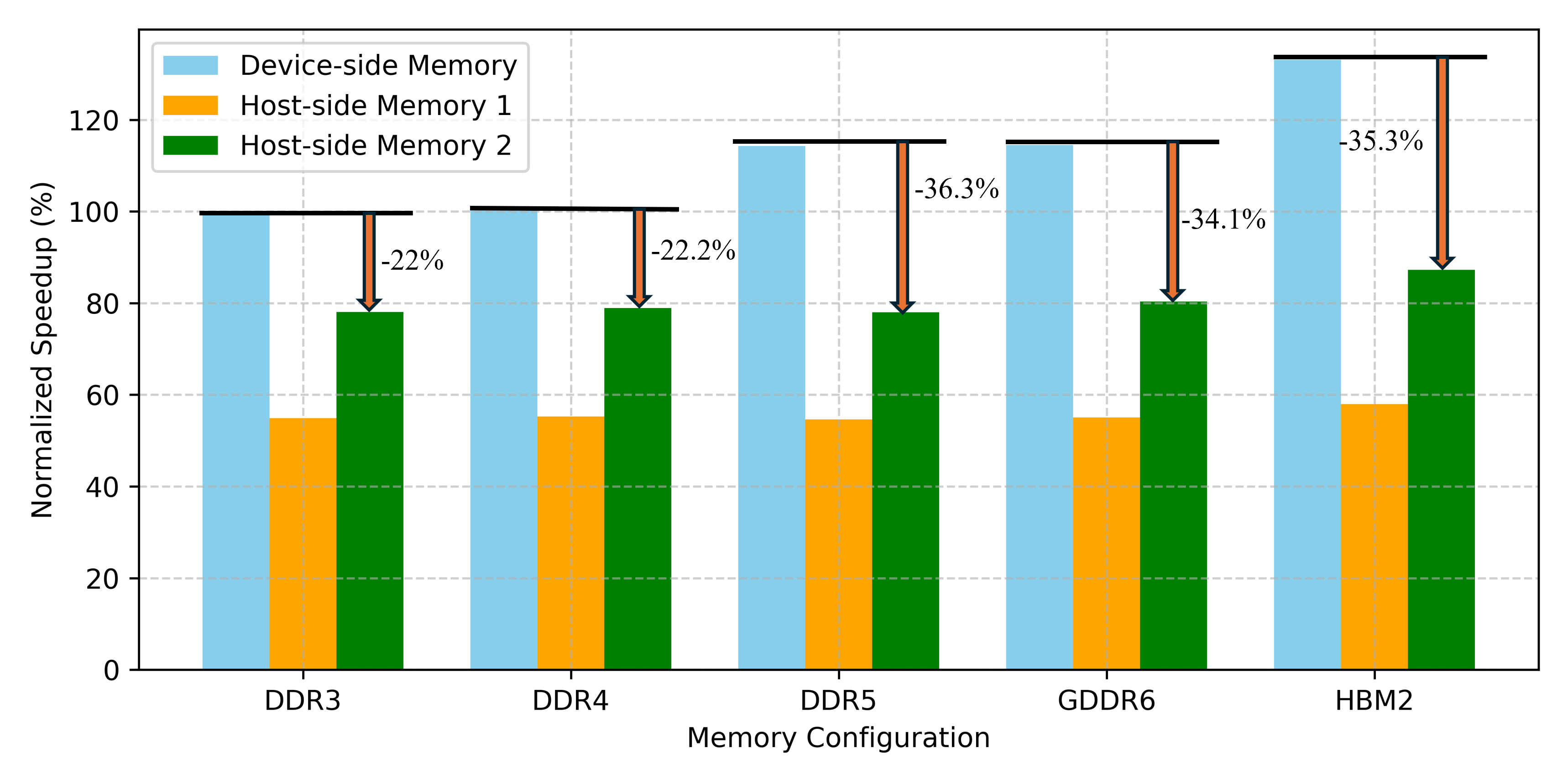}
  \vspace{-0.5cm}
  \caption{Impact of DRAM type and location}
  \vspace{-0.25cm}
  \label{fig:device_vs_host_memory}
\end{figure}

Fig.~\ref{fig:device_vs_host_memory} presents the normalized speedup (wrt DDR4 device-side data) comparison between device-side memory and two host-side memory configurations (one with a 2GB/s and one with a 64GB/s PCIe link) for DDR4, HBM, GDDR5, and LPDDR5. 

The results demonstrate that device-side memory consistently outperforms host-side memory across all tested memory types, regardless of the speed of the PCIe bus. Host-side memory shows lower speedups, with a clear dependence on PCIe speed. When using the 64GB/s PCIe configuration, host-side memory can achieve around 78\% of the performance relative to device-side memory, which is a substantial improvement over the 2 GB/s PCIe. Performance degradation depends on the memory technology, and is aggravated for GDDR5 and HBM, as the gap between host-side and device-side memory grows larger.


\textbf{\textit{Key Takeaway \#3:} Implementing device-side memory significantly boosts performance, improving accelerator efficiency by up to 2 times compared to all other configurations.}


\paragraph{Memory bandwidth and latency sweeping}
We investigate the impact of memory bandwidth and latency by varying one while keeping the other one constant, using HBM as a case study with gem5's default DRAM model.

\vspace{-0.25cm}
\begin{figure}[h]
\vspace{-0.25cm}
  \centering
  \includegraphics[width=\linewidth]{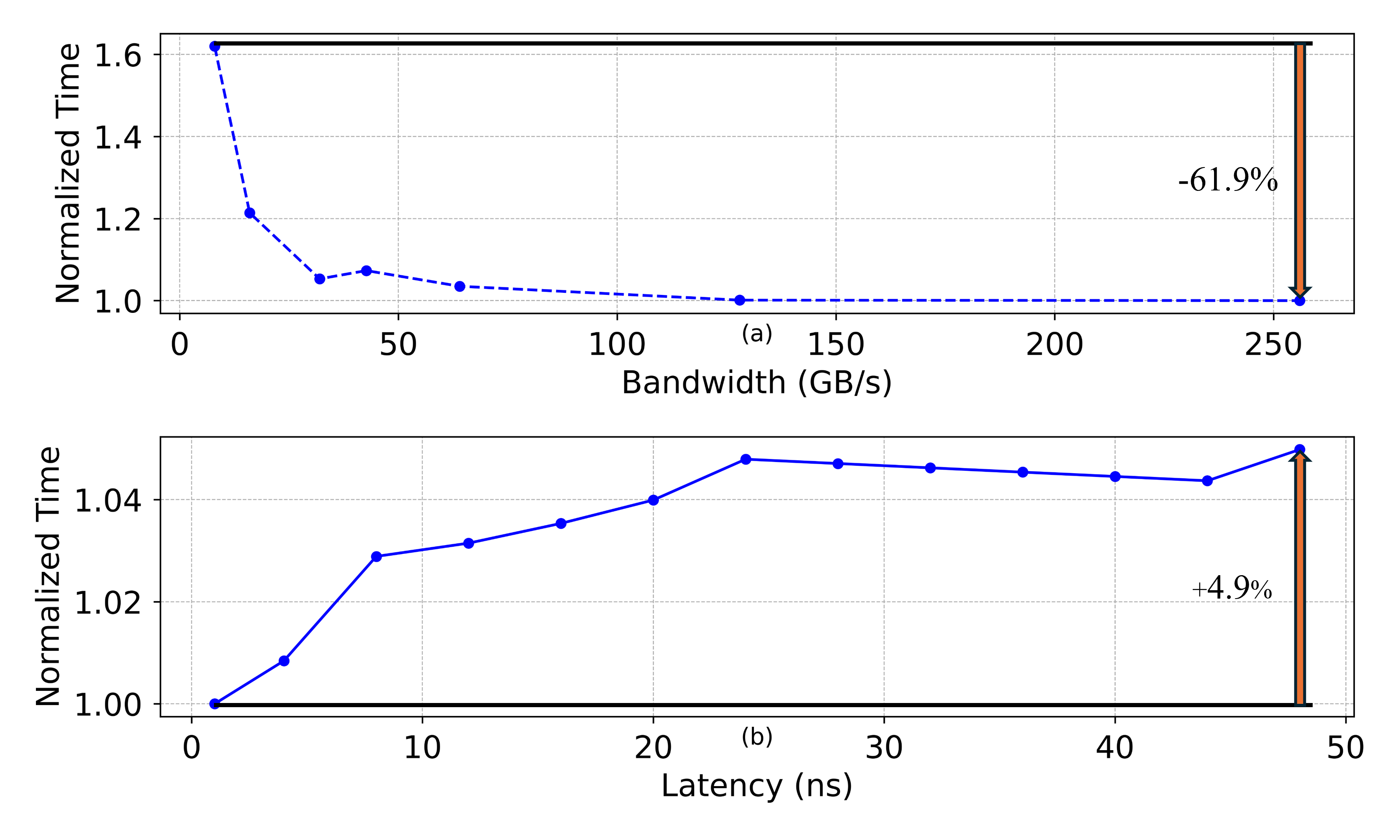}
  \vspace{-0.6cm}
  \caption{Impact of Memory Bandwidth(a) and Memory Latency(b)}
  \label{fig:Performance_Bandwidth}
\end{figure}
\vspace{-0.25cm}

Fig.~\ref{fig:Performance_Bandwidth}(a) shows that normalized execution time decreases significantly as bandwidth increases up to approximately 50GB/s, yielding up to a 60\% performance improvement. This underscores the critical role of bandwidth in alleviating performance bottlenecks at lower levels. Beyond 100GB/s, the curve plateaus, with only a 1.7\% improvement from 50GB/s to 256GB/s, thus other system components become the bottleneck.


Fig.~\ref{fig:Performance_Bandwidth}(b)  illustrates that execution time increases with latency, especially sharply from 1ns to 12ns, then plateaus between 12ns and 36ns. However, the overall time overhead due to increased latency is only about 4.9\%. This analysis indicates that while both bandwidth and latency affect performance, the system is significantly more sensitive to changes in bandwidth. Thus, bandwidth is the more crucial factor in GEMM-dominated workloads.

\textbf{\textit{Key Takeaway \#4:} Bandwidth has a greater impact than latency in GEMM performance: bandwidth improves performance by 60\% and latency only adds 5\% overhead.}

\subsubsection{Address Translation performance analysis}

In this section, we analyze the impact of virtual address translation on system performance, focusing on larger matrix sizes. Table~\ref{tab:addrtrans} summarizes key metrics such as memory footprint, translation times, and page table walk (PTW) times.

As matrix size increases, translation overhead initially decreases, reaching 1.00\% at 1024, but rises to 6.49\% at 2048. This reflects how larger matrices reduce per-translation costs through amortization but face increasing complexity with larger address spaces. Similarly, the mean translation time drops to its lowest value at 1024 but spikes to 54.38 cycles at 2048, indicating inefficiencies. PTW times also increase with matrix size, reaching 368.14 cycles at 2048, highlighting the growing complexity of address translation for very large datasets. 
This highlights the critical need for careful optimization of address translation mechanisms to maintain system performance, particularly for large-scale workloads.

\begin{table*}[htbp]
\vspace{-0.25cm}
\centering
\caption{Results with Larger Matrix Sizes}
\label{tab:addrtrans}
\begin{tabular}{lcccccc}
\toprule
\textbf{Metric} & \textbf{64} & \textbf{128} & \textbf{256} & \textbf{512} & \textbf{1024} & \textbf{2048} \\
\midrule
Memory Footprint (Pages) & 12.0 & 48.0 & 192.0 & 768.0 & 3072.0 & 12288.0 \\
Translation Times & 3130 & 18470 & 142738 & 1082780 & 8593259 & 68430699 \\
Trans Mean Time & 23.42683 & 20.37948 & 13.87159 & 9.91735 & 10.478634 & 54.38005 \\
PTW Times & 15 & 54 & 227 & 1034 & 7675 & 479244 \\
PTW Mean Time & 176.6666 & 281.90740 & 265.255507 & 252.465184 & 294.609381 & 368.141137 \\
uTLB Lookup times & 2350 & 17690 & 137290 & 1081610 & 8586250 & 68423690 \\
uTLB Misses times & 195 & 862 & 8644 & 65808 & 731513 & 10416279 \\
Trans Overhead & 6.02\% & 1.87\% & 1.59\% & 1.30\% & 1.00\% & 6.49\% \\
\bottomrule
\end{tabular}
\vspace{-0.5cm}
\end{table*}

\textbf{\textit{Key Takeaway \#5:} Address translation overhead significantly impacts performance for large memory footprints. Larger matrices increase translation complexity and page table walk times, requiring optimization to avoid bottlenecks.}
\vspace{-0.25cm}
\subsection{Transformer Performance Evaluation}
We apply the GEMM analysis insights to Transformer inference using four system configurations:
\begin{enumerate}
    \item A system using host memory and small PCIe bandwidth of 2 GB/s with 4 4Gbps lanes (PCIe-2GB).
    \item A system using host memory and moderate PCIe bandwidth of 8 GB/s with 8 8Gbps lanes (PCIe-8GB).
    \item A system using host memory and PCIe bandwidth of 64 GB/s with 16 64Gbps lanes (PCIe-64GB).
    \item A system without host memory, utilizing device-side memory instead (DevMem).
\end{enumerate}

We select ViT as the target workload and configure each system based on the GEMM analysis:
\begin{itemize}
    \item For PCIe-2GB and PCIe-8GB: packet size of 256 and DDR4 memory.
    \item For PCIe-64GB: packet size of 256 and HBM2 memory.
    \item For DevMem: packet size of 64 and HBM2 memory.
\end{itemize}
\vspace{-0.25cm}
\begin{figure}[h]
\vspace{-0.25cm}
    \centering
    \includegraphics[width=\linewidth]{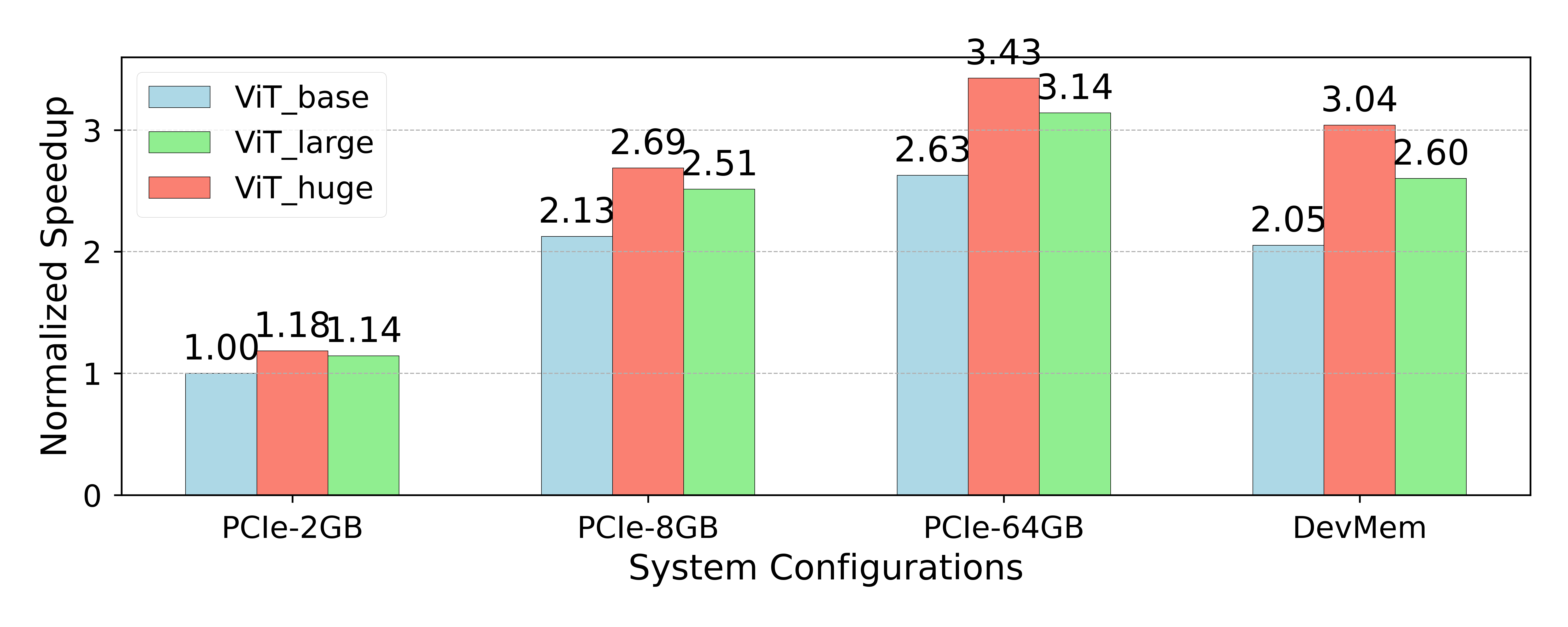}
    \vspace{-0.8cm}
    \caption{Performance comparison of memory locations and interconnects}
    \label{fig:config_comparison}
    \vspace{-0.25cm}
\end{figure}

Fig.~\ref{fig:config_comparison} shows that PCIe-64GB achieves significant performance improvements over the baseline PCIe-2GB, with speedups ranging from approximately 2.5x to 3.4x, highlighting the critical impact of PCIe bandwidth when using host memory. In contrast, the DevMem configuration, despite leveraging device-side memory (HBM2 with burst size of 64) and reduced data transfer times due to proximity to computational units, performs slightly worse than PCIe-64GB.
\vspace{-0.1cm}
\subsection{Transformer Workload Analysis}
\subsubsection{GEMM and Non-GEMM Performance Evaluation}
\begin{figure}[h]
    \centering
    \includegraphics[width=\linewidth]{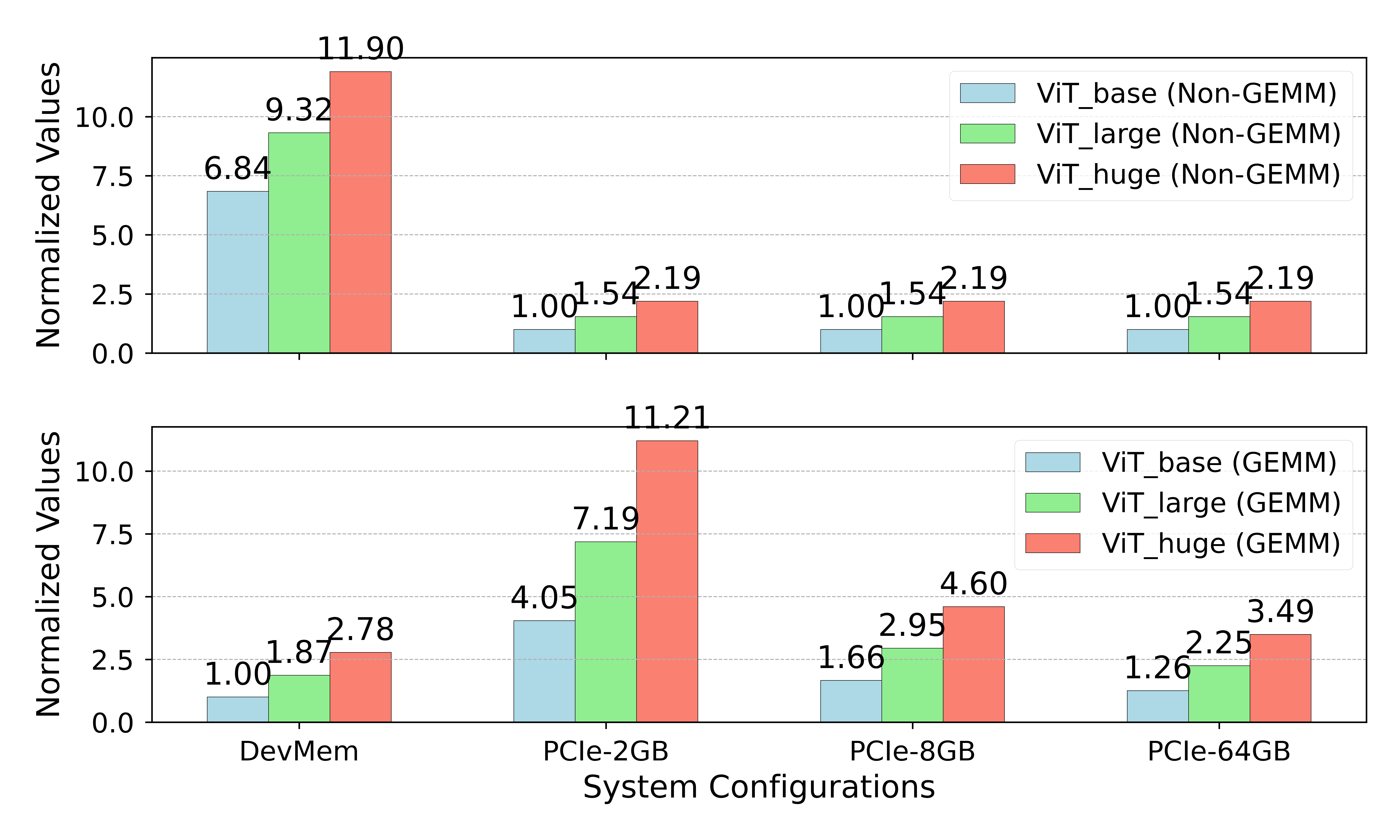}
    \vspace{-0.4cm}
    \caption{Performance Comparison of GEMM and Non-GEMM workload}
    \label{fig:NonGEMM_comparison}
    \vspace{-0.2cm}
\end{figure}

To understand the performance discrepancies observed, we conduct a detailed analysis of the Transformer workload, profiling both GEMM and Non-GEMM operations across different ViT models and system configurations, as highlighted in prior research~\cite{Ivanov2020},~\cite{Karami2024}. Fig.~\ref{fig:NonGEMM_comparison} shows that while DevMem offers the best performance for GEMM workloads due to its high data bandwidth, it exhibits the poorest performance in Non-GEMM operations, incurring up to a 500\% overhead compared to systems using the PCIe interface. This degradation stems from the high latency introduced by the NUMA architecture and longer access times to device-side memory. Consequently, despite DevMem's advantages in GEMM tasks, these overheads hinder system performance in Non-GEMM computations.

\textbf{\textit{Key Takeaway \#6:} Device-side memory overhead from NUMA architecture can degrade performance, accounting for ~40\% in Transformer experiments like ViT\_large.}

\subsubsection{Performance Analysis of GEMM and Non-GEMM Workloads}
To quantify the performance trade-offs between GEMM and Non-GEMM workloads, we propose a model that expresses the total execution time of Transformers as a combination of these two components. By understanding their respective contributions and computational requirements, we can determine the optimal balance between device-side and host-side memory utilization.
The total execution time \( \text{Time}_{\text{overall}} \) of a Transformer workload can be expressed as:
\vspace{-0.25cm}
\[
\text{Time}_{\text{overall}} = T_{\text{other}} + \frac{W_{\text{GEMM}}}{P_{\text{GEMM}}} + \frac{W_{\text{NonGEMM}}}{P_{\text{NonGEMM}}}
\]

Where \( T_{\text{other}} \) is the fixed time for other operations.
\( W_{\text{GEMM}} \) is the fraction of GEMM workload (\( 0 \leq W_{\text{GEMM}} \leq 1 \)).
\( P_{\text{GEMM}} \) and \( P_{\text{NonGEMM}} \) are the performance metrics for GEMM and Non-GEMM workloads, respectively. In our case, the Non-GEMM percentage represents the proportion of overall time spent on Non-GEMM workloads when executed on a PCIe system configuration.

\begin{figure}[h]
    \centering
    \includegraphics[width=\linewidth]{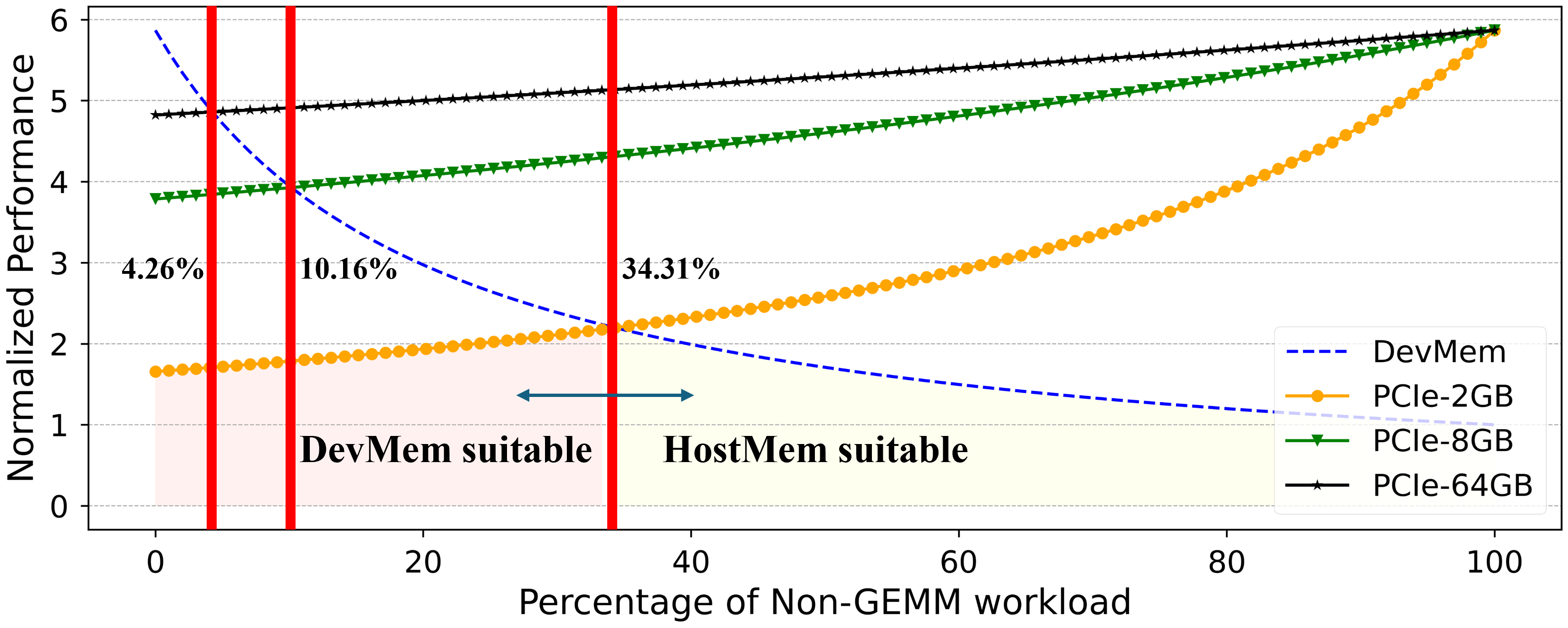}
    \vspace{-0.8cm}
    \caption{Overall Transformer Performance as a Function of Non-GEMM Workload Fraction for various PCIe Bandwidths vs DevMem}
    \label{fig:NonGEMMPer_comparison}
    \vspace{-0.5cm}
\end{figure}

We vary the \( W_{\text{NonGEMM}} \) from 0\% to 100\% to analyze the performance variance. Fig.~\ref{fig:NonGEMMPer_comparison} shows the performance variation with increasing Non-GEMM percentage. Our analysis reveals that DevMem outperforms the PCIe system when the GEMM workload fraction (\( W_{\text{GEMM}} \)) exceeds certain thresholds, which decrease as PCIe bandwidth increases. Specifically, DevMem is preferable when \( W_{\text{GEMM}} \) exceeds 34.31\% for a PCIe bandwidth of 2\,GB/s, 10.16\% for 8\,GB/s, and 4.27\% for 64\,GB/s. This indicates that as PCIe bandwidth increases, the advantage of using DevMem diminishes unless the workload is overwhelmingly dominated by GEMM operations.

\textbf{\textit{Key Takeaway \#7:} The choice between PCIe and DevMem depends on workload composition (GEMM vs. Non-GEMM) and PCIe bandwidth; DevMem is preferred when Non-GEMM fractions are below a threshold.}

\vspace{-0.15cm}
\section{Conclusion}
\vspace{-0.15cm}
In this paper, we have introduced Gem5-AcceSys, a comprehensive framework that extends the Gem5 simulator to support standard interconnects like PCIe, NUMA architectures, and configurable memory hierarchies. This advancement addresses critical limitations of existing simulators, enabling detailed system-level co-design and realistic performance evaluation of hardware accelerators. Using Gem5-AcceSys, we conduct an in-depth analysis of a matrix multiplication accelerator tailored for transformer workloads. Our experiments demonstrate that optimized PCIe interconnects can achieve up to 80\% of the performance of systems using device-side memory, and in some cases, even surpass them. We also uncover trade-offs between GEMM and Non-GEMM workloads, highlighting how their balance influences the optimal choice of memory configurations and interconnect strategies. By facilitating the exploration of standard interconnects and memory hierarchies, Gem5-AcceSys provides actionable insights for system architects, empowering them to balance performance and cost in designing efficient and scalable next-generation accelerator systems.

\vspace{-0.15cm}
\section*{Acknowledgments}
\vspace{-0.15cm}
We thank the anonymous reviewers for their feedback, Pengbo Yu for hardware and manuscript guidance, and Gabriel Catel Torres and Clément Dieperink for software support. 

This work was supported in part by the Swiss State Secretariat for Education, Research, and Innovation (SERI) through the SwissChips research project, and also by Intel as part of the Intel Center for Heterogeneous Integrated Platforms (HIP).

\vspace{12pt}

\end{document}